\documentclass[12pt]{article}
\begin{document}

\title{Nonstandard order parameters\\and the origin of CP violation}
\author{\normalsize B. Holdom\\\small {\em Department of Physics,
University of Toronto}\\\small {\em Toronto, Ontario,} M5S1A7,
CANADA}\date{}\maketitle
\begin{picture}(0,0)(0,0)
\put(310,205){UTPT-99-12}
\end{picture}

\begin{abstract} The consideration of chirality-preserving 2-fermion order
parameters may shed new light on the strong CP problem and the
breakdown of flavor symmetries. We describe two situations, one having
the standard KM picture for weak CP violation and another having new
sources of weak CP violation.
\end{abstract}
\baselineskip 19pt

With the discovery/confirmation \cite{dd} of a nonvanishing \(\varepsilon
'\), it appears likely that the Kobayashi-Maskawa (KM) picture for weak CP
violation will be confirmed at the \(B\)-factories. There remains one fly in
the ointment, though, and that is the strong CP problem. The invisible axion
scenario
\cite{gg} is not completely satisfactory, since it relates the smallness of the
strong CP violating parameter \(\overline{\theta }\) to a poorly understood
hierarchy of symmetry breaking scales. Any alternate resolution of the
problem will likely involve phases entering the mass matrix in a very
particular way
\cite{hh}, or in a very limited way
\cite{ii}, in order to satisfy the constraint on
\(\mathit{arg}\mathit{det}M\). This restricts the phase structure of the KM
matrix and typically necessitates nonstandard sources of weak CP
violation. This has been the focus of recent work \cite{jj}. A common
feature of these alternate approaches is the central role played by
elementary scalar fields, since CP is required to be broken softly in the
scalar field sector before feeding into the quark sector.

We wish to consider new possibilities for the origin of CP violation which
may arise in the context of dynamical symmetry breaking, in the absence of
elementary scalar fields. We consider possible order parameters
constructed from the quark fields themselves, which may form due to the
participation of the standard quarks in strong flavor interactions on roughly
1000 TeV scales. The latter is a typical flavor scale needed to account for
the light quark masses. The manner in which the phases are transmitted to
the quark mass matrix and other observables can differ substantially from
models with scalar fields. We are motivated by the strong CP problem, but
we will also describe the resulting picture for weak CP violation. We will
discuss how one choice of the order parameters can lead to the standard
KM picture, and how another choice implies that new sources of CP
violation are likely to exist.

We are exploring the possibility that various flavor symmetries and CP are
both violated dynamically on the scale of strong flavor interactions. Above
these scales we assume that there is nothing but a CP conserving, massless
gauge theory. Naively a dynamical breakdown of CP by strong interactions
would sprinkle CP violating phases everywhere in the low energy theory,
including the effective operators responsible for the quark masses, and
result in a \(\overline{\theta }\) much too large. We will suggest that the
proliferation of CP violating phases can be constrained by the pattern of
symmetry breaking.

In particular we suppose that CP violation originates in the phases of a pair
of order parameters which are also responsible for breaking some
\(\mathit{U}(1)\) flavor symmetry. We will be considering mass matrices
such that some of the elements respect a \(\mathit{U}(1)\) symmetry while
other elements do not. We will describe how the former elements are
protected from CP violating phases, while the latter pick up a phase from
one of the CP violating order parameters. At the end we will discuss how
the \(\mathit{U}(1)\) can be part of a \textit{gauged} flavor symmetry.

Before describing the order parameters we first consider three possible sets
of mass matrices, each having a vanishing \(\mathit{arg}\mathit{det}M\).
In each case we also give the charge of the flavor symmetry we are
considering, where \({N_{i}}\) is the \(i\)th family number. Each \#
denotes in general a different real number.
\begin{eqnarray}
\lefteqn{\mathrm{Case\ 1}{\ \ \ \ }{Q_{F}}={N_{1}} + {N_{2}} -
{N_{3}}} \nonumber\\
 & & M^{U}= \left(\!   {\begin{array}{ccc}
\# & \# & \#e^{{ - i{\phi _{U}}}} \\
\# & \# & \#e^{{ - i{\phi _{U}}}} \\
\#e^{{i{\phi _{U}}}} & \#e^{{i{\phi _{U}}}} & \#
\end{array}}
 \!\right)  {\ \ \ \ }M^{\mathit{D}}= \left(\!   {\begin{array}{ccc}
\# & \# & \#e^{{ - i{\phi _{\mathit{D}}}}} \\
\# & \# & \#e^{{ - i{\phi _{\mathit{D}}}}} \\
\#e^{{i{\phi _{\mathit{D}}}}} & \#e^{{i{\phi _{\mathit{D}}}}} & \#
\end{array}}
 \!\right)   {\label{g}}\end{eqnarray}\pagebreak
\begin{eqnarray}
\lefteqn{\mathrm{Case\ 2}{\ \ \ \ }{Q_{F}}= - {N_{1}} + {N_{2}} +
{N_{3}}} \nonumber\\
 & & M^{U}= \left(\!   {\begin{array}{ccc}
\# & \#e^{{i{\phi _{U}}}} & \#e^{{i{\phi _{U}}}} \\
\#e^{{ - i{\phi _{U}}}} & \# & \# \\
\#e^{{ - i{\phi _{U}}}} & \# & \#
\end{array}}
 \!\right)  {\ \ \ \ }M^{\mathit{D}}= \left(\!   {\begin{array}{ccc}
\# & \#e^{{i{\phi _{\mathit{D}}}}} & \#e^{{i{\phi _{\mathit{D}}}}} \\
\#e^{{ - i{\phi _{\mathit{D}}}}} & \# & \# \\
\#e^{{ - i{\phi _{\mathit{D}}}}} & \# & \#
\end{array}}
 \!\right)  \end{eqnarray}
\begin{eqnarray}
\lefteqn{\mathrm{Case\ 3}{\ \ \ \ }{Q_{F}}={N_{1}} - {N_{2}} +
{N_{3}}} \nonumber\\
 & & M^{U}= \left(\!   {\begin{array}{ccc}
\# & \#e^{{i{\phi _{U}}}} & \# \\
\#e^{{ - i{\phi _{U}}}} & \# & \#e^{{ - i{\phi _{U}}}} \\
\# & \#e^{{i{\phi _{U}}}} & \#
\end{array}}
 \!\right)  {\ \ \ \ }M^{\mathit{D}}= \left(\!   {\begin{array}{ccc}
\# & \#e^{{i{\phi _{\mathit{D}}}}} & \# \\
\#e^{{ - i{\phi _{\mathit{D}}}}} & \# & \#e^{{ - i{\phi 
_{\mathit{D}}}}} \\
\# & \#e^{{i{\phi _{\mathit{D}}}}} & \#
\end{array}}
 \!\right)   {\label{i}}\end{eqnarray} The key property of these matrices is
that they are form invariant: addition and multiplication of matrices of the
form of \(M^{U}\) will again yield matrices of the same form with the
same phase, and the same for \(M^{\mathit{D}}\). This is necessary since
sums of mass-type diagrams involving multiple insertions of mass matrices
(and/or flavor changing order parameters) are expected to contribute to the
final mass matrix. Although the phases appear in a rather trivial way, for
\({\phi _{U}} - {\phi _{\mathit{D}}}\neq n\pi \) we will have a CP
violating phase appearing in the KM matrix.

We would now like to identify the flavor-violating order parameters which
could introduce phases into the mass matrices in this way. First, the order
parameters must respect electroweak symmetry since we are supposing that
they arise at the \(\approx 1000\) TeV flavor scale. Second, to guarantee a
unique phase in each mass matrix, it is clear that there must not be two or
more order parameters in either the up- or down-sectors having independent
phases. We might expect for example two order parameters with opposite
\(\mathit{U}(1)\) flavor charge, but we must insist that these be hermitian
conjugates of each other so that their phases are equal and opposite.

An order parameter with these properties is one which is not normally
considered, namely the chirality-preserving part of the quark propagator. In
particular we suppose that the origin of CP violation lies in the following
amplitudes coupling \textit{right-handed} quarks of \textit{different}
families. We assume that the pattern of symmetry breaking produces these
order parameters for \textit{only one pair} of families labeled by indices
\(({f_{1}}, {f_{2}})\), with \({f_{1}}\neq {f_{2}}\), which are such as to
break the \(\mathit{U}(1)\) flavor symmetry.
\begin{equation} {\displaystyle \int } e^{{ip\cdot x}}\langle \mathrm{T}
{U_{{f_{1}}}}(x){\overline{U}_{{f_{2}}}}(0)\rangle dx= ie^{{i{\phi
_{U}}}}{\displaystyle \frac {1
 + {\gamma _{5}}}{2}}  Z^{U}(p^{2}){\displaystyle \frac
{1}{{/\!\!\!}p}} {\label{a}}\end{equation}
\begin{equation} {\displaystyle \int } e^{{ip\cdot x}}\langle \mathrm{T}
{\mathit{D}_{{f_{1}}}}(x){\overline{\mathit{D}}_{{f_{2}}}}(0)
\rangle dx=ie^{{i{\phi _{\mathit{D}}}}}{\displaystyle \frac {1 +
{\gamma _{5}}}{ 2}} Z^{\mathit{D}}(p^{2 }){\displaystyle \frac
{1}{{/\!\!\!}p}} {\label{b}}\end{equation} The \(\mathit{U}(1)\)
symmetry implies that \({\phi _{U}} + {\phi _{\mathit{D}}}\) is freely
adjustable, leaving only \({\phi _{U}} - {\phi _{\mathit{D}}}\) as a
physical phase. Note that our symmetry is vectorial; the corresponding
purely right-handed transformation is not a symmetry of the rest of the
theory. If there were no weak interactions, and no other physics caused
transitions between the up and down sectors, then there would be two
vector \(\mathit{U}(1)\) symmetries and then no physical phase would
remain. This emphasizes the role of weak interactions in producing CP
violation of any kind in this picture.

We expect other independent order parameters generated by the strong
flavor interactions, which signal the breakdown of various other
symmetries. For example there may be 4-fermion order parameters, since
they can also respect electroweak symmetry, but which break other flavor
and/or chiral symmetries. Such 4-fermion order parameters may play a
crucial role in generating the quark masses (in the presence of some
fermions with TeV scale masses) \cite{cc}.

Our central assumption is that the order parameters in (\ref{a},\ref{b}) are
the \textit{only}
\textit{source} of flavor \(\mathit{U}(1)\) and CP breaking. To clarify what
we mean by this we note that these order parameters are nonperturbative
contributions to the full quark propagators. We can imagine rewriting the
theory in terms of the full propagators; this can take the form of the CJT
\cite{ee} effective action or some other similar formalism. In this alternate
form of the theory, in which the breakdown of the flavor and CP
symmetries are encoded in the Feynman rules, we are supposing that the
effects of flavor \(\mathit{U}(1)\) and CP symmetry breaking can be
described \textit{entirely perturbatively}. (This may be taken as a
statement about the size of the effective coupling in this alternate
description.) In this way we are drawing a distinction between the
dynamical breakdown of flavor and CP as reflected by the order
parameters, and the subsequent appearance of flavor and CP breaking in
other amplitudes of the theory.

 \(\mathit{U}(1)\) breaking and CP violation in a general amplitude can
then be deduced from the effects of all possible insertions of the order
parameters. Ignoring weak corrections for the moment, the result is that
\textit{the CP violating phase is correlated with the flavor charge of the
amplitude}. This is evident in the mass matrices (\ref{g}-\ref{i}). Consider
also the chirality-preserving terms for the right-handed quark fields, which
in the low energy theory are of the form\pagebreak
\begin{equation} {\overline{U}_{Rj}}i{\,/\!\!\!\!}
\mathit{D}{C_{\mathit{jk}}^U}{U_{Rk}} +
{\overline{U}_{Rj}}i{\,/\!\!\!\!}
\mathit{D}\mathit{D^2}{C_{\mathit{jk}}^{{\prime }U}}{U_{Rk}} +
{...}{\label{e}}\end{equation}
\begin{equation} {\overline{\mathit{D}}_{Rj}}i{\,/\!\!\!\!
}\mathit{D}{C_{\mathit{jk}}^\mathit{D}}{\mathit{D }_{Rk}} +
{\overline{\mathit{D}}_{Rj}}i{\,/\!\!\!\!}\mathit{D}\mathit{D^2}{C_{\mathit{jk}}^{{\prime
}D}}{\mathit{D}_{Rk}} + {...} {\label{f}}\end{equation} The Hermitian
matrices \(C^{U}\) and \(C^{\mathit{D}}\), and the matrices in the
nonrenormalizable terms, will have the same phase structure as the mass
matrices \(M^{U}\) and \(M^{\mathit{D}}\) respectively. That is, they
have the same phases in the same elements.

In the low energy theory one can transform the renormalizable kinetic
terms to conventional flavor-diagonal form by redefining the fields,
\begin{equation} {U_{R}}\rightarrow H^{U}{U_{R}}{\ \ \ \ }, {\ \ \ \
}{\mathit{D}_{R}}
\rightarrow H^{\mathit{D}}{\mathit{D}_{R}}.\end{equation} The
\(H^{U}\) and \(H^{\mathit{D}}\) matrices can be chosen to be Hermitian
and they again have the same phase structure as the original \(M^{U}\) and
\(M^{\mathit{D}}\) respectively. Under this transformation the mass
matrices change, but they retain their original phase structure. Thus in the
low energy theory we have quarks with normal kinetic terms, and mass
matrices with a CP violating phase which preserves \(\overline{\theta
}=0\). The order parameters in (\ref{a},\ref{b}) are affecting the mass
matrices both through direct insertions in mass generating diagrams and
through the field redefinition we have just described.

This discussion has ignored weak interaction corrections, and in fact these
generate the well known small contributions to \(\overline{\theta }\)
\cite{bb}. In addition to a safely small finite contribution, the standard
model generates ``infinite'' contributions occurring at high orders in the
weak coupling. In our picture such effects simply represent the running of
\(\overline{\theta }\) below the flavor scale; above the flavor scale
\(\overline{\theta }\) is trivially zero if the fundamental theory is a CP
conserving, massless gauge theory. It is the softness of the dynamically
generated quantities at the flavor scale, as in (\ref{a},\ref{b}), which turns
the ``infinite'' \(\overline{\theta }\) of the standard model into a calculable
and safely small quantity.

Any new physics which, like the weak interactions, couples together the up
and down sectors would have to be carefully considered as a source of
additional contributions to \(\overline{\theta }\). One example is
\({\mathit{SU}(2)_{R}}\) interactions. Another would be nonperturbative
operators of the form
\({\overline{Q}_{\mathit{Lj}}}{Q_{\mathit{Rk}}}{\overline{Q}_{\mathit{Lk}}}{Q_{\mathit{Rj}}}\)
where \(j\) and \(k\) denote different families. These operators could
connect an off-diagonal element of the up mass matrix to one in the down
mass matrix, and thus could produce contributions to both mass matrices of
a form different than we have been considering. We assume that such
effects are either nonexistent or sufficiently small.

The flavor physics will generate nonrenormalizable terms suppressed by
the large flavor scale, such as those in (\ref{e},\ref{f}), but these do not
lead to corrections which affect the phase structure of the mass matrices.
They may however produce contributions to weak CP observables, in
addition to the phase in the KM matrix; we will return to such effects later.

We will now consider the matrices presented in (\ref{g}-\ref{i}) in more
detail. Case 1 turns out to be of the most interest if our goal is to obtain a
standard KM picture for weak CP violation. This may also be possible in
Case 2, but we will explain why it is more likely to have additional sources
of CP violation in this case. In Case 3, for any choice of \(({f_{1}},
{f_{2}})\), we are not able to obtain realistic masses and mixings, and we
will not consider this case further.

For realistic mass matrices in Case 1 we are forced to choose \(({f_{1}},
{f_{2}})=(2, 3)\), so that transitions occur between \({U_{2R}}\) and
\({U_{3R}}\) and between \({\mathit{D}_{2R}}\) and
\({\mathit{D}_{3R}}\) in the original flavor basis. For the resulting mass
terms \(({\overline{Q}_{\mathit{Lj}}}{M_{\mathit{jk}}^U}
{U_{\mathit{Rk}}} +
{\overline{Q}_{\mathit{Lj}}}{M_{\mathit{jk}}^\mathit{D}}{\mathit{D}_{\mathit{Rk}}})
+ \mathrm{h.c.}
\) we specialize somewhat from the matrices in (\ref{g}) and consider the
following, where all parameters are real.
\begin{equation} M^{U}= \left(\!   {\begin{array}{ccc}{m_{11}^U} &
{m_{12}^U} & e^{{ - i{\phi _{U}}}}{\rho _{U}}{m_{12}^U}
\\{m_{21}^U} & {m_{22}^U} & e^{{ - i{\phi _{U}}}}{\rho
_{U}}{m_{22}^U} \\ 0 & e^{{i{\phi _{U}}}}{\rho _{U}}{m_{33}^U}
& {m_{33}^U }
\end{array}}
 \!\right)  \end{equation}
\begin{equation} M^{\mathit{D}}= \left(\!  
{\begin{array}{ccc}{m_{11}^\mathit{D}} & {m_{12}^\mathit{D}} &
e^{{ - i{\phi  _{\mathit{D}}}}}{\rho
_{\mathit{D}}}{m_{12}^\mathit{D}} \\{m_{21}^\mathit{D}} &
{m_{22}^\mathit{D}} & e^{{ - i{\phi  _{\mathit{D}}}}}{\rho
_{\mathit{D}}}{m_{22}^\mathit{D}} \\ 0 & e^{{i{\phi
_{\mathit{D}}}}}{\rho _{\mathit{D}}}{m_{33}^\mathit{D}} &
{m_{33}^\mathit{D}}
\end{array}}
 \!\right)  \end{equation} We have used a common proportionality constant
\({\rho _{U}}\) in \(M^{U}\), and \({\rho _{\mathit{D}}}\) in
\(M^{\mathit{D}}\), although this need not be strictly true. This choice
reflects how these elements can arise from the neighboring mass elements
in the same row, via the right-handed transitions. Zeros have been placed in
the \((3, 1)\) elements since they are likely to be small and unimportant to
the results; they would arise from diagrams involving multiple insertions of
the mass matrices.

By studying the transformation from the present weak eigenstate basis to
the mass eigenstate basis we can deduce properties of the KM matrix
\(V\). We find for example that, up to small corrections,
\begin{equation}
 \left|   {V_{\mathit{cb}}}   \right| \approx  \left|    {X_{\mathit{D}}} -
e^{(i({\phi _{\mathit{D}}} - {\phi _{U}}))} {X_{U}}   \right|
,\end{equation} where
\begin{equation} {X_{U}}=2{\displaystyle \frac {{m_{c}}}{{m_{t}}}}
{\displaystyle \frac {{\rho _{U}}}{1 - {\rho _{U}^2}}} {\ \ \ }, {\ \ \ }{X
_{\mathit{D}}}=2{\displaystyle \frac {{m_{s}}}{{m_{b}}}}
{\displaystyle \frac {{\rho  _{\mathit{D}}}}{1 - {\rho
_{\mathit{D}}^2}}} .\end{equation} In addition we find that the bulk of
the contribution to \( \left|   {V_{\mathit{cb}}}   \right| \) must come from
\({X_{U}}\) (so as to obtain realistic values for  \( \left|  
{V_{\mathit{ub}}}   \right| \)) which then implies that \({\rho
_{U}}\approx .9\). Thus the effects of the right-handed transitions must be
large. For two of the angles of the unitarity triangle we find
\begin{equation}
\alpha \approx {\phi _{U}} - {\phi _{\mathit{D}}}, {\ \ \ \ \ \ \  }\gamma
\approx \arg({X_{\mathit{D}}} - e^{(i({\phi _{\mathit{D}}} - {\phi
_{U}}))}{X_{U}}).\end{equation} The possibility that \({\phi _{U}} -
{\phi _{\mathit{D}}}=\pi /2\) (corresponding to a relative factor of \(i\)
between the order parameters in (\ref{a},\ref{b})) is certainly consistent
with a realistic \(\alpha \).

We illustrate these results with the following values.
\begin{equation} {m_{11}^U}=.002, {m_{12}^U}={m_{21}^U}=0,
{m_{22}^U}=4.4, {m_{33}^U}=117\end{equation}
\begin{equation} {m_{11}^\mathit{D}}=0,
{m_{12}^\mathit{D}}={m_{21}^\mathit{D}}=.019,
{m_{22}^\mathit{D}}=.08, {m_{33}^\mathit{D}}=2.7\end{equation}
\begin{equation} {\rho _{U}}=.9, {\rho _{\mathit{D}}}=.25, {\phi
_{U}}=\pi / 2, {\phi _{\mathit{D}}}=0\end{equation} This yields mass
eigenvalues \(({m_{u}}, {m_{c}}, {m_{t}})=(.002,  .62, 158)\) and
\(({m_{d}}, {m_{s}}, {m_{b}})=\)\linebreak \((.0043, .077, 2.8)\), which
are typical of realistic values renormalized at the TeV scale. The unitarity
triangle has the following angles and sides.
\begin{equation}
\alpha =88.6^{{\circ}}{\ \ \ \ },  {\ \ \ \ }\beta =21.9^{{\circ}} {\ \ \ \ }, {\ \ \
\ }\gamma =69.5^{{\circ}}\end{equation}
\begin{equation} 1:{R_{b}}:{R_{t}}=1: .37:.94\end{equation} (Note that
in this example with \({m_{12}^U}={m_{21}^U}=0\) the origin of
Cabbibo mixing lies in the down sector. We can get similar final results
with Cabbibo mixing originating in the up sector, which would require that
\({m_{12}^U}\gg {m_{21}^U}\).) It is of interest to consider the KM
matrix \(V=L^{{U{\dagger }}}L^{\mathit{D}}\) using a phase convention
in which the unitary transformations \(L^{U}\) and \(L^{\mathit{D}}\)
have the same phase structure as \(M^{U}\) and \(M^{\mathit{D}}\)
respectively. The phases in \(V\)  then occur mostly in the \(2\times 2\)
block involving the heavier quarks:
\begin{equation} V= \left(\!   {\begin{array}{ccc} .976 & .218 & .00331 \\
 - .218 - {.000007}i & .975 - .000536i
 & .0139 + .0374i \\
 - .000196 - .00814i &  - .0143 + .0365i & .999 + .000521i
\end{array}}
 \!\right)  \end{equation} Note that \({V_{\mathit{ud}}}\) and
\({V_{\mathit{us}}}\) are always real, no matter what the value of \({\phi
_{U}}\) and \({\phi _{\mathit{D}}}\). The next case we consider will be
different.

For Case 2 we are forced to choose \(({f_{1}}, {f_{2}})=(1, 2)\), which
leads to the following matrices.
\begin{equation} M^{U}= \left(\!   {\begin{array}{ccc}{m_{11}^U} &
e^{{i{\phi _{U}}}}{\rho _{U}}{m_{11}^U} &  0 \\ e^{{ - i{\phi
_{U}}}}{\rho _{U}}{m_{22}^U} & {m_{22}^U}
 & {m_{23}^U} \\ e^{{ - i{\phi _{U}}}}{\rho _{U}}{m_{32}^U} &
{m_{32}^U}
 & {m_{33}^U}
\end{array}}
 \!\right)  \end{equation}
\begin{equation} M^{\mathit{D}}= \left(\!  
{\begin{array}{ccc}{m_{11}^\mathit{D}} & e^{{i{\phi
_{\mathit{D}}}}}{\rho _{\mathit{D}}}{m_{11}^\mathit{D}} & 0 \\ e^{{
- i{\phi _{\mathit{D}}}}}{\rho _{\mathit{D}}}{m_{22}^\mathit{D}} &
{m_{22}^\mathit{D}} & {m_{23}^\mathit{D}} \\ e^{{ - i{\phi
_{\mathit{D}}}}}{\rho _{\mathit{D}}}{m_{32}^\mathit{D}} &
{m_{32}^\mathit{D}} & {m_{33}^\mathit{D}}
\end{array}}
 \!\right)  \end{equation} For Cabibbo mixing we find
\begin{equation}
 \left|   {V_{\mathit{us}}}   \right| \approx  \left|    {Y_{\mathit{D}}} -
e^{(i({\phi _{\mathit{D}}} - {\phi _{U}}))} {Y_{U}}   \right|
.\end{equation} where
\begin{equation} {Y_{U}}=2{\displaystyle \frac {{m_{u}}}{{m_{c}}}}
{\displaystyle \frac {{\rho _{U}}}{1 - {\rho _{U}^2}}} {\ \ \ }, {\ \ \ }{Y
_{\mathit{D}}}=2{\displaystyle \frac {{m_{d}}}{{m_{s}}}}
{\displaystyle \frac {{\rho  _{\mathit{D}}}}{1 - {\rho
_{\mathit{D}}^2}}} .\end{equation} The \({Y_{\mathit{D}}}\) term
dominates and thus \({\rho _{\mathit{D}}}\) is determined. For the angle
\(\beta \) appearing in the unitarity triangle we find
\begin{equation}
\beta \approx \arg({Y_{\mathit{D}}} - e^{(i({\phi _{\mathit{D}}} - {\phi
_{U}}))}{Y_{U}}).\end{equation} If \({\rho _{U}}\) is much smaller
than .95 the unitarity triangle is very thin, even if \({\phi _{U}} - {\phi
_{\mathit{D}}}\approx \pi /2\), and it is then not possible for the KM
matrix to account for the observed CP violation. Independent of this we are
able to obtain the other three real mixing parameters of the KM matrix.

We illustrate the case of a fat unitarity triangle with the following values.
\begin{equation} {m_{11}^U}=-.03, {m_{11}^U}=-.035,
{m_{23}^U}={m_{32}^U}=8.1, {m_{33}^U}=157\end{equation}
\begin{equation} {m_{11}^\mathit{D}}=-.0133,
{m_{11}^\mathit{D}}=-.035, {m
_{23}^\mathit{D}}={m_{32}^\mathit{D}}=.26,
{m_{33}^\mathit{D}}=2.75\end{equation}
\begin{equation} {\rho _{U}}=.95, {\rho _{\mathit{D}}}=.76, {\phi
_{U}}=\pi  /2, {\phi _{\mathit{D}}}=0\end{equation} This yields mass
eigenvalues \(({m_{u}}, {m_{c}}, {m_{t}})=(.0021,  .62, 158)\) and
\(({m_{d}}, {m_{s}}, {m_{b}})=(.0044,  .076, 2.8)\). The unitarity
triangle has following angles and sides.\pagebreak
\begin{equation}
\alpha =75.0^{{\circ}}{\ \ \ \ },  {\ \ \ \ }\beta =16.8^{{\circ}} {\ \ \ \ }, {\ \ \
\ }\gamma =88.2^{{\circ}}\end{equation}
\begin{equation} 1:{R_{b}}:{R_{t}}=1: .30:1.03\end{equation} Note that
in contrast to Case 1 (with the same phase convention as used there) the
phases in the KM matrix now appear mostly in the \(2\times 2\) block
involving the lighter quarks.
\begin{equation} V= \left(\!   {\begin{array}{ccc} .975 + .0140i & .211 -
.0645i &  - .000667 - .00262i \\
 - .211 - .0646i & .974 - .0140i & .0398 + .000044i \\ .00910 + .000018i & 
- .0388 + {.000004}i & .999
\end{array}}
 \!\right)  {\label{h}}\end{equation}
\indent In the standard model it is conventional to absorb phases into the
quark fields so as to move phases in \(V\) into more standard positions, and
in particular remove phases from the \({V_{\mathit{ud}}}\) and
\({V_{\mathit{us}}}\) elements. But in our case the new flavor physics
generates additional nonrenormalizable operators which are not invariant
under these phase redefinitions. In particular let us consider the flavor
changing chromomagnetic moment operators in the down sector, which
have been considered as a new physics contribution to \(\varepsilon
'/\varepsilon \) \cite{aa},
\begin{equation}
{g_{s}}{{\tilde{C}}_{\mathit{jk}}^\mathit{D}}{\overline{\mathit{D}}_{Lj}}\sigma
^{{\mu \nu }}T^{a}{\mathit{D}_{Rk}}{G_{\mu \nu
}^a}.{\label{c}}\end{equation} The origin of these operators should be
closely associated with the origin of the down-type quark masses. Thus
very roughly \({{\tilde{C}}_{\mathit{jk}}^\mathit{D}}\approx
\sqrt{{m_{i}}{m_{j}}}/\Lambda ^{2}\), where \(\Lambda \) should be of
order the electroweak symmetry breaking scale. Although
\({\tilde{C}^\mathit{D}}\) will have the same phase structure as
\(M^{\mathit{D}}\), there is no reason for \({\tilde{C}^\mathit{D}}\) to be
exactly proportional to \(M^{\mathit{D}}\), and in the basis in which
\(M^{\mathit{D}}\) is diagonal we expect \({\tilde{C}^\mathit{D}}\) to
have nondiagonal entries. With the quark phase choice \({\phi
_{\mathit{D}}}=0\), \({\tilde{C}^\mathit{D}}\) is real, while the phases in
the KM matrix are nonstandard. We can use phase redefinitions on the
quark fields (which leave the diagonal mass matrix and \(\overline{\theta
}\) invariant) to move the phases in the KM matrix to more standard
positions. Then an explicit CP violating phase will show up in
\({{\tilde{C}}_{\mathit{ds}}^\mathit{D}}\). The magnitude of this phase
is \( \left|   \arg({V_{\mathit{us}}})   \right| \) in the original basis, and
this, as it turns out, is closely approximated by \(\beta \). Thus the same
angle which determines the extent of CP violation in the KM matrix also
determines the CP violating phase in the new operator.

We may estimate the \(\Lambda \) required for this effect to account for the
observed \(\varepsilon '/\varepsilon \) \cite{dd}. Assuming that \( \left|  
\mathrm{Im} ({{\tilde{C}}_{\mathit{ds}}^\mathit{D}})
   \right| \approx \sqrt{{m_{d}}{m_{s}}}\beta /\Lambda ^{2 }\) we deduce
from the analysis in \cite{aa} that \(\Lambda /\sqrt{\beta }\approx 5\) TeV.
Since this is above the electroweak breaking scale, this new contribution to
\(\varepsilon '\) can easily be large enough, even for values of \(\beta \)
which make the KM contribution tiny.

To account for \(\varepsilon \) itself, we need a superweak, \(\Delta S=2\),
4-fermion operator. Such operators emerge naturally from the exchange of
new flavor gauge bosons, which can lead for example to a  \(\frac
{1}{M^{2}}({\overline{\mathit{D}}_{1R}}{\gamma _{\mu
}}{\mathit{D}_{1R}} - {\overline{\mathit{D}}_{2R}}{\gamma _{\mu
}}{\mathit{D}_{2R}} )^{2}\) operator in the original flavor basis (\(M\)
characterizes the flavor scale). In the mass eigenstate basis this will
produce a \(\Delta S=2\) operator, which then picks up a phase due to the
phase redefinitions we have just discussed. Accounting for large mixing
between \({\mathit{D}_{1R}}\) and \({\mathit{D}_{2R}}\), due to a large
\({\rho _{\mathit{D}}}\), we obtain roughly \(M/\sqrt{\beta }\approx
10^{4}\) TeV. Since this is above the expected scale of new flavor physics,
the superweak contribution to \(\varepsilon \) can easily be large enough,
even for values of \(\beta \) which make the KM contribution tiny.

It we look back at Case 1 we see that both \({V_{\mathit{uu}}}\) and
\({V_{\mathit{us}}}\) are real in the original flavor basis, and thus the
interference between the new operators and standard model amplitudes
introduces no new phases. Thus the new contributions to \(\varepsilon \)
and \(\varepsilon '\) occurring in Case 2 do not occur in Case 1. For Case 2,
although it is conceivable for the KM matrix by itself to account for CP
violation, we have seen that it may be more natural to have the bulk of CP
violation originate in additional operators.

Let us step back for a moment and consider the larger picture. Quark
masses are generated by 4-fermion interactions in the presence of some
dynamically generated TeV scale fermion masses. It is natural to assume
that these TeV scale fermions are nothing but a fourth family of quarks
\((\mathit{t'}, \mathit{b'})\) and leptons, in which case the possible origin
of the \(\mathit{U}(1)\) flavor symmetry we have been discussing becomes
clearer. By extending the \(\mathit{U}(1)\) flavor symmetry to the fourth
family it can become weak anomaly free and thus can correspond to part of
some gauged flavor symmetry. For example the \(\mathit{U}(1)\) flavor
charge in Case 1 becomes \({N_{1}} + {N_{2}} - {N_{3}} - {N_{4}}\),
which is light family number minus heavy family number. This can
correspond to the diagonal generator of a \(\mathit{SU}(2)\) flavor
symmetry which acts on two pairs of families, such that the flavor
eigenstates for quarks \(({Q_{1}}, {Q_{3}})\) and \(({Q_{2}},
{Q_{4}})\) are flavor doublets. Alternatively if \(({Q_{1}}, {Q_{3}})\)
transforms as the complex conjugate of \(({Q_{2}}, {Q_{4}})\) under
\(\mathit{SU}(2)\) then the diagonal \(\mathit{SU}(2)\) generator is \( -
{N_{1}} + {N_{2}} + {N_{3}} - {N_{4}}\), and this gives us Case 2.

The four-fermion interactions respecting the \(\mathit{U}(1)\) would
generate mass matrices of the following form (with the ordering
\(({Q_{1}}, {Q_{2}}, {Q_{3}}, {Q_{4}})\)), for Case 1 and Case 2
respectively,
\begin{equation}
 \left(\!   {\begin{array}{cccc}
\# & \# & 0 & 0 \\
\# & \# & 0 & 0 \\ 0 & 0 & \# & \# \\ 0 & 0 & \# & \#
\end{array}}
 \!\right)  {\ \ \ \ }, {\ \ \ \ } \left(\!   {\begin{array}{cccc}
\# & 0 & 0 & \# \\ 0 & \# & \# & 0 \\ 0 & \# & \# & 0 \\
\# & 0 & 0 & \#
\end{array}}
 \!\right)  .{\label{d}}\end{equation} To this we must add the effects of the
right-handed transitions, which involve the \(({f_{1}}, {f_{2}})=(2, 3)\)
flavors in Case 1 and the \(({f_{1}}, {f_{2}})=(1, 2)\) flavors in Case 2. In
both cases it is natural for the mixing between the fourth family and the
lighter families to be small, in which case our previous discussion will
continue to hold.

Finally, suppose that the fourth family mass also breaks the
\(\mathit{U}(1)\) flavor symmetry. This would occur in the case of
complex conjugate \(\mathit{SU}(2)\) representations if the fourth family
quark mass term was \({\overline{Q}_{4L}}{Q_{3R}} + \mathrm{h.c}\).
(Then of course the \({Q_{3}}\) and \({Q_{4}}\) families in the original
flavor basis are no longer close to being the mass eigenstates.) By
reordering the right-handed fields, \(({Q_{1R}}, {Q_{2R}}, {Q_{3R}},
{Q_{4R}} )\rightarrow ({Q_{2R}}, {Q_{1R}}, {Q_{4R }}, {Q_{3R}})\),
it then turns out that the mass matrix takes the same form as the second
matrix in (\ref{d}). We only mention this case here to make contact with
the model in
\cite{cc}.\footnote{Note that in that reference various symmetry breaking
effects in the quark sector were postulated to feed in from the fourth family
lepton sector. In the present discussion those contributions are not needed,
due to the chirality-preserving order parameters.}

In conclusion we have related CP violation to a phase mismatch in certain
flavor changing order parameters involving right-handed quarks. The
absence of strong CP violation is related to the very particular way these
order parameters feed phases into the mass matrices. We described two
cases, one with a standard KM picture of weak CP violation, and the other
where additional sources of CP violation in the \(K\) system are likely. In
the latter case there may be smaller than expected CP violation in the \(B\)
system. In both cases we related the angles of the unitarity triangle to the
phases in the flavor changing order parameters. We cannot claim a final
resolution of the strong CP problem without a complete and unambiguous
theory of quark masses, and thus we await the results from the \(B\)
factories for further guidance.

\section*{Acknowledgment} This research was supported in part by the
Natural Sciences and Engineering Research Council of Canada. I thank T.
Torma for his comments on the manuscript.

\end{document}